# On the Interaction between Fermions via Mass less Bosons and Massive Particles

Voicu Dolocan[1], Voicu Octavian Dolocan[2] and Andrei Dolocan[3*]

[1]*Faculty of Physics, University of Bucharest, Bucharest, Romania*
[2]*Aix-Marseille University & IM2NP, Avenue Escadrille Normandie Niemen, 13397 Marseille cedex 20, France*
[3]*National Institute for Laser, Plasma and Radiation Physics, Bucharest, Romania*

**Abstract:** By using a Hamiltonian based on the coupling through flux lines, we have calculated the interaction energy between two fermions via massless bosons as well as via massive particles. In the case of interaction via massless bosons we obtain an equivalent expression for the Coulomb's energy on the form $\alpha \hbar c / r$, where $\alpha$ is the fine structure constant. In the case of the interaction via massive particles we obtain that the interaction energy contains a term building the potential well. Taking into account the spin-spin interaction of the nucleons, we show that this interaction modulates the interaction potential through a cosine factor. The obtained results are in good agreement with experimental data, for example, of deuteron.

## 1. Introduction

It is assumed that there are four fundamental interactions in nature: the electromagnetic force, strong interaction, weak interaction and gravitation. The strength of the strong interaction is 100 times that of the electromagnetic force, and several orders of magnitude greater than that of the weak force and gravitation. These ratios are in fact figure of convenience; the fundamental figures vary dramatically according to the distances over which the forces are exerted. For example, the weak force is as strong as the electromagnetic force at the distances relevant to its effects, which are a small fraction of the radius of the nucleus of an atom. The strong interaction is observable in two ranges: On the smaller scale it is the force that holds quarks and gluons together to form the proton, the neutron and other particles. On the larger scale, it is also the force that binds protons and neutrons together to form the nucleus of an atom. In the context of binding protons and neutrons (nucleons) together to form atoms, the strong interaction is called the nuclear force (or residual strong force). In this case it is the residuum of the strong interaction between the quarks that make up the protons and neutrons. As such, the residual strong interaction obeys a quite different distance-dependent behavior between nucleons, than when it is acting to bind quarks within nucleons. The strong force is thought to be mediated by gluons, acting upon quarks, antiquarks, and the gluons themselves. This is detailed in the theory of quantum chromodynamics [1,2].

In this paper we present the potential energy of interaction between fermions via massless bosons as well as via massive particles, by using a Hamiltonian based on the coupling between two bodies through flux lines[3,4]. We take into consideration the spin-spin interaction of the nucleons and apply our results to the deuteron.

## 2. The Energy of Interaction via Massless Bosons

We have obtained [3,4] that the energy of interaction between two fermions via bosons is given by the following expression (see Appendix A):

$$E_{int} = \frac{\hbar^3 D^2}{32 M^2 r^2 \left(\rho_o + \frac{Dr}{c^2}\right)^2} \sum_{q,q_o,k} \frac{(q \cdot q_o)^2}{\omega_q^2 \omega_{q_o}^2} \frac{1}{2} \left|\sum_n e^{iq_o \cdot r_n}\right|^2 \frac{\omega_q}{(\varepsilon_k - \varepsilon_{k-q})^2 - \omega_q^2} \quad (1)$$

where $D$ is a coupling constant, M is the mass of a fermion, $r$ is the distance between the two fermions, $\rho_o$ is the massive density of the interacting field [6,7], $Dr/c^2$ is the "massless density" of the interacting field, $\omega_q = cq$ is the classical oscillation frequency of the interacting field, $\omega_{q_o}$ is the oscillation frequency of a fermion, $q$ is the wave vector of the interacting field, $q_o$ is the wave vector of the boson associated with the electron, $k$ is the wave vector of the electron and $\varepsilon_k = \frac{\hbar k^2}{2M}$. In the case $\rho_o = 0$, for a quasi free fermion, when $\omega_{q_o} = \hbar q_o^2 / 2M, \varepsilon_k - \varepsilon_{k-q} \ll \omega_q$, may be written:

$$\sum_q \frac{(q \cdot q_0)^2}{\omega_q^3 \omega_{q_0}^2} = \left(\frac{2M}{\hbar}\right)^2 \frac{1}{q_o^2 c^3} \frac{\Omega}{(2\pi)^2} \int_0^\pi \cos^2 \alpha \sin \alpha d\alpha \int_0^{q_o} q dq = \left(\frac{2M}{\hbar}\right)^2 \frac{r^3}{9\pi c^3}$$

and

$$2 \sum_{q_o} [1 + \cos(q_o \cdot r)] = 2 + 2 \frac{\Omega}{(2\pi)^2} \int_0^{0.76\pi/r} q_o^2 dq_o \int_0^\pi \cos(q_o r \cos\theta) \sin\theta d\theta = 3.3$$

We have considered $\sum_k 1 = 1$ and $\Omega = 4\pi r^3 / 3$. The upper limit of $q_0$ as $0.76\pi/r$ is imposed by the constraint condition $[(4\pi r^3 / 3)/(2\pi)^3] \times 4\pi q_o^3 / 3 = 1$. These results are also valid for $k = 0$.

The interaction energy (1) becomes

$$E_{int} = -\frac{1}{147} \frac{\hbar c}{r} \approx -\frac{\alpha \hbar c}{r} \quad (2)$$

where $\alpha = 1/147$ differs some few from the fine structure constant. This is an equivalent expression for the Coulomb energy of interaction. In fact the fine structure constant is $1/137 = 0.0073$, the value which may be obtained if the upper limit over $q_o$ will be equal to $0.94\pi/r$, a value with 3.2% larger than that obtained from the constraint condition. This means that the boundaries of the wave function of the system are not steep. If instead of Fröhlich fraction[5]

$$\frac{\omega_q}{(\varepsilon_k - \varepsilon_{k-q})^2 - \omega_q^2} \quad (3a)$$

we use the fraction

$$\frac{-1}{(\varepsilon_k - \varepsilon_{k-q}) - \omega_q} \quad (3b)$$

the sign in front of the relation (2) is positive. The relation (3a) is written when one of the interacting particles emits and the other absorbs a field boson, and may be assumed that in this case the two interacting particles have opposite sign. The relation (3b) is written when the two interacting particles absorbs field bosons and may be assumed that they have the same sign, so that there is a repulsion.

We have found this expression without taken into account the concept of the electric charge. We specify that neither the charge nor the mass of the electron or any charged particle can actually by calculated in quantum electrodynamics- they have to be assumed.

### 3. The Energy of Interaction via Massive Particles

Now we assume that $\Box(\varepsilon_\mathbf{k} - \varepsilon_\mathbf{k-q}) = mc^2$ and by using the above relations, Eq. (1) where the last fraction is substituted by the fraction (3b), becomes

$$E_{int} = -\frac{\hbar c}{16 r^4} \sum_{q,q_o,k} \frac{\cos^2\alpha}{q_o^2} |\sum_n e^{iq_o \cdot r_n}|^2 \frac{1}{\left(\frac{mc}{\hbar}\right) - q} \qquad (4)$$

Further,

$$\sum_q \frac{\cos^2\alpha}{\left(\frac{mc}{\hbar}\right) - q} = \frac{\Omega}{(2\pi)^3} 2\pi \int_0^\pi \cos^2\alpha \sin\alpha d\alpha \int_0^{q_o} \frac{q^2 dq}{\left(\frac{mc}{\hbar}\right) - q} =$$

$$\frac{2}{3} \frac{\Omega}{(2\pi)^2} \left(\frac{mc}{\hbar}\right)^2 \int_0^{(\hbar/mc)q_o} \frac{x^2 dx}{1-x} = \frac{2\Omega}{3(2\pi)^2} \left(\frac{mc}{\hbar}\right)^2 \left\{-\frac{\hbar}{mc} q_o - \frac{1}{2}\left(\frac{\hbar}{mc}\right)^2 q_o^2 - \ln|1 - \left(\frac{\hbar}{mc}\right) q_o|\right\}$$

and

$$\sum_{q_o,q} \frac{\cos^2\alpha}{q_o^2} |\sum_n e^{iq_o \cdot r_n}|^2 \frac{1}{\left(\frac{mc}{\hbar}\right) - q} =$$

$$\frac{2}{3}\frac{\Omega}{(2\pi)^2}\left(\frac{mc}{\hbar}\right)^2 \left\{-\frac{2\hbar}{mc}\frac{\Omega}{(2\pi)^2}\int_0^{0.76\pi/r} q_o dq_o \int_0^\pi [1+\cos(q_o r\cos\theta)]\sin\theta d\theta - \left(\frac{\hbar}{mc}\right)^2 \left[1 + \frac{\Omega}{(2\pi)^2}\int_0^{0.76\pi/r} q_o^2 dq_o \int_0^\pi \sin\theta d\theta\right] - \right.$$

$$\left. \frac{4\Omega}{(2\pi)^2} \int_0^{0.76\pi/r} \log|1 - \frac{\hbar}{mc} q_o| - \frac{2\Omega}{(2\pi)^2} \int_0^{0.76\pi/r} dq_o \log|1 - \frac{\hbar}{mc} q_o| \int_0^\pi \cos(q_o r\cos\theta)\sin\theta d\theta\right\}$$
(5)

Finally, one obtains

$$\frac{E_{int}}{mc^2} = 0.0087 + \frac{0.0068}{x} + 0.0019 x^2 \int_0^{2.42/x} dz \left[1 + \frac{\sin(xz)}{xz}\right] \log|1-z| \qquad (6)$$

where $x = rmc/\hbar$ and m is the mass of the exchanged particle between the two interacting fermions.

The first term, $0.0087 mc^2$, is the kinetic energy. The second term is the repulsive Coulomb energy of interaction, which is $(0.0068/x)mc^2 \approx (1/147)\hbar c/r$, and represents the electromagnetic energy of interaction. The last term represent the energy of interaction due to exchange of massive particles [6,7]

$$v_N = 0.0019 x^2 \int dz \left[1 + \frac{\sin(xz)}{xz}\right] \log|1-z| \qquad (6a)$$

In Fig. 1 are represented the reduced nuclear potential energy, $v_N$, as well as, the reduced interaction energy $E = E_{int}'/mc^2$ as a function of the reduced distance between the two fermions. The plot is calculated in Scilab with a step size of $\Delta x = 0.02$. In Table I ( Appendix B), we present the numerical data. It is observed that the reduced nuclear potential energy presents a minimum of -0.018644131520 at x = 2.18, while the total interaction energy E presents a minimum of 0.006851838985 at $x$ = 2.2. If we subtract from this energy, the contribution of the kinetic term and we write $V_c = E - 0.0087$, then the depth of the potential well is equal to $\Delta V_c$ = -0.015551838986. Next, we compare these results with experimental data for deuteron. The deuteron is the simplest of all the nucleon bound states i.e., the atomic nuclei. It is therefore, particularly, suitable for studying the nucleon-nucleon interaction. The proton-neutron system is mostly made up of an $l$ = 0 state. The experimental value of the binding energy of the deuteron is B = 2.225 MeV. According to virial theorem the binding energy B is equal to minus (1/2) from the potential energy, that is $B = -\Delta V_c \times mc^2/2$.

We can evaluate the mass of a particle exchanged by the two interacting nucleons from the relation

$$m = -\frac{B}{(\Delta V_c) c^2} \qquad (7)$$

By using the data presented above one obtains $m \approx$ 287 MeV/$c^2$. This value is close to the experimental value of 300 MeV/$c^2$. The proton has two $u$ (up)- quark and one $d$ (down)-quark, while the neutron has two $d$ – quarks and one $u$ – quark. Our results show that the interaction between a proton and a neutron occurs as follows: the proton emits a $u$ – quark and absorbs a $d$ – quark, while the neutron emits a $d$ – quark and absorbs a $u$ – quark. Therefore, permanently, the proton and the neutron change their positions. The minimum of the potential well is x=2.2, so that

$$r = (\hbar/mc) x = 1.4 \text{ fm}$$

which is in the range of the nuclear force. The potential energy of interaction, $V_c$ contains in addition to $v_N$, the repulsive Coulomb energy. The question is: we must take out the repulsive Coulomb contribution in the case of the neutron-proton interaction? If the answer is yes, then the potential energy of the deuteron is $v_N$(6a).

## 4. Effects of the Spin-Spin Interactions

Now we will take into account the interaction between the spins of the two interacting nucleons. The natural unit for expressing magnetic dipole moment of heavy particles such as nucleons and atomic nuclei is the nuclear magneton, which is defined by

$$\mu_N = \frac{e\hbar}{2M_p c}$$

where $M_p$ is the proton rest mass. The magnetic moment of the proton is $\mu_p = +2.79267\mu_N$ and the magnetic moment of the neutron is $\mu_n = -1.913139\,\mu_N$. We define the potential vector as:

$$A = \frac{\mu \times r}{R^3}$$

and we substitute in the above equations $q_o$ by

$$q_o' = q_o - \frac{e}{\hbar c}\frac{\mu \times r}{r^3}$$

and we write in Eq. (5)

$$\left|\sum_{n=1,2} e^{iq'r_{on}\cdot r_n}\right|^2 = 2(1+\cos\Gamma)$$

$$\Gamma = q_2\cdot r_2 - q_1\cdot r_1 - \frac{e}{\hbar c}\left[\oint_{l_2} A(r)dr - \int_{l_1} A(r)dr\right] = \frac{1}{2}(q_2+q_1)(r_2-r_1) +$$

$$\frac{1}{2}(q_2-q_1)(r_2+r_1) - \Gamma_o = q_o r\cos\theta - \Gamma_o$$

$$\Gamma_o = \frac{e}{\hbar c}\left[\oint_{l_2}\frac{\mu_p \times r}{r^3}d\cdot r - \oint_{l_1}\frac{\mu_n ¿ \times r}{r^3}d\cdot r\right]$$

where we have considered $q_1 = q_2 = q_o$ and $r_2 - r_1 = r$. By using the expressions

$$\mu_p = g_p\mu_N\sigma_p, \quad \mu_n = g_n\mu_N\sigma_n$$

one obtains

$$\Gamma_o = \frac{e}{\hbar}\frac{e\hbar}{2Mc^2}[g_p\oint(\frac{\sigma_p \times r}{r^3})d\cdot r - g_n\oint(\frac{\sigma_n \times r}{r^3})d\cdot r] = ¿$$

$$\frac{e^2}{2Mc^2}[\frac{2\pi \times 2.79267}{r} + \frac{2\pi \times 1.913139}{r}] = \frac{4.7058\,\pi e^2}{Mrc^2} \quad (8a)$$

when $\sigma_n \;\uparrow\downarrow\; \sigma_p,$ and

$$\Gamma'_o = \frac{0.86531\,\pi\,e^2}{Mrc^2} \tag{8b}$$

when $\boldsymbol{\sigma}_n \uparrow\uparrow \boldsymbol{\sigma}_p$; $\boldsymbol{\sigma}_n$, $\boldsymbol{\sigma}_p$ are the spin unit vectors for the two nucleons. We have considered, $M_p = M_n = M$ and $\boldsymbol{\sigma}_p$, $\boldsymbol{\sigma}_n$ are the spin unit vectors for the two nucleons and $g_p$, $g_n$ are $g$ nuclear factors, respectively. In this situation, instead of relation (6) one obtains

$$\frac{E_{int}(\Gamma_o)}{mc^2} = 0.0054 + 0.0033\cos\left(\frac{0.034122}{x}\right) + \frac{0.0045}{x} + \frac{0.0023}{x}\cos\left(\frac{0.034122}{x}\right) + \\ 0.0019\,x^2 \int_0^{2.42/x} dz \ln|z-1|\left\{1 + \frac{\sin(xz)}{xz}\cos\left(\frac{0.034122}{x}\right)\right\} \tag{9}$$

for the case (8a), and an analogous relation for the case (8b) where instead of $\Gamma_o = 0.03412/x$, we have $\Gamma''_o = 0.00633331/x$. We have used the result of the following integral

$$\int_0^\pi \cos(q_o r\cos\theta - \Gamma_o)\sin\theta d\theta = 2\cos(\Gamma_o)\frac{\sin(q_o r)}{q_o r}$$

It is observed that the spin-spin interaction modifies the interaction potential through a cosine factor. The first two terms from the right hand of expression (9) represents the kinetic energy, the next two terms represents the Coulomb energy of interaction, and the last term is due to the interaction via massive particles. In Table I are presented numerical data for $E_{int}(\Gamma_o)/mc^2$ as well as for $E_{int}(\Gamma'_o)/mc^2$. The minimum in the first case is equal to -0.006851517954, while in the second case is equal to -0.006851930681. The minima are situated at x = 2.2. The difference in energy between the two states of the deuteron is $\Delta E$ = 92 eV. So the deuteron pass from a parallel spin state for the two nucleons to an antiparallel spin state, or vice-versa, by emitting and absorbing an energy quanta of ~ 90 eV. This energy quanta may increase or decrease its kinetic energy.

## 5. Solution of the Deuteron with Central and Tensor Potential

One writes the Schrödinger equation for the deuteron system as[8-16]:

$$-\frac{\hbar^2}{M_p}\nabla^2 \Psi_M(r) - \left[E' - V_c(r) - V_T(r)S_{12}\right]\Psi_M(r) = 0 \\ S_{12} = \frac{3(\sigma_1 \cdot r)(\sigma_2 \cdot r)}{r^2} - \sigma_1 \cdot \sigma_2 \tag{10}$$

$V_c(r)$ is the central potential given by Eq. (6)

$$V_c(r) = mc^2 v_c(x) \\ v_c(x) = \frac{0.0068}{x} + 0.0019\,x^2 \int_0^{2.42/x} dz\left[1 + \frac{\sin(xz)}{xz}\right]\ln|z-1| \tag{11a}$$

and $V_T(r)S_{12}$ is the tensor potential, $V_T(r)$ is given by Eq. (9),.

$$V_T(r) = mc^2 v_T(x)$$

$$v_T(x) = \frac{0.0045}{x} + \frac{0.0023}{x}\cos\left(\frac{0.034122}{x}\right) + 0.0019 x^2 \int_0^{2.42/x} dz \ln|z-1|\left\{1 + \frac{\sin(xz)}{xz}\cos\left(\frac{0.034122}{x}\right)\right\}$$

(11b)

. We will assume $\Psi_M(r)$ as a mixture of $L=0$ and $L=2$ only. It is assumed $M_n \approx M_p$. Under these conditions we can write the wave function for the ground state of the deuteron as

$$\Psi_M(r) = \frac{u(r)}{r} Y_{01}^M(\theta,\varphi) + \frac{w(r)}{r} Y_{121}^M(\theta,\varphi) \qquad (12)$$

where

$$Y_{JLS}^M(\theta,\phi) = \sum_{M_L, M_S} \langle J, M | L, M_L; S, M_S \rangle Y_{L,M_L}(\theta,\phi)\chi_{S,M_S}$$

$$M = M_L + M_S$$

where $Y_{L,M_L}$ and $\chi_{S,M_S}$ are spherical harmonics and spin wave function respectively, $\mathbf{S} = (1/2)(\sigma_1 + \sigma_2)$. The reduced radial wave functions $u(r)$ and $w(r)$ correspond to S and D waves respectively. Following standard procedure[8-14] one obtains the two coupled equations

$$\frac{\hbar^2}{M_p}\frac{d^2 u}{dr^2} - \left[V_c(r) - E'\right]u = \sqrt{8} V_T(r) w$$

$$\frac{\hbar^2}{M_p}\left[\frac{d^2 w}{dr^2} - \frac{6w}{r^2}\right] - \left[V_c(r) - 2V_T(r) - E'\right]w = \sqrt{8} V_T(r) u$$

(13)

The discussion of these equations is difficult since no exact solutions exist in terms of tabulated functions. So a numerical calculus is necessary. In order to determine the unknown parameters are added the following relations. From the wave function, a characteristic size of the deuteron $r_m$ is defined as the rms-half distance between the two nucleons

$$r_m^2 = \frac{1}{4}\int_0^\infty \left[u^2(r) + w^2(r)\right] r^2 dr \qquad (14)$$

The deuteron electric quadrupole moment is also determined from the wave functions

$$Q_d = \frac{1}{\sqrt{50}}\int_0^\infty r^2 uw\, dr - \frac{1}{20}\int_0^\infty r^2 w^2 dr \qquad (15)$$

In this simple potential model of the deuteron, the magnetic moment of the deuteron is determined entirely by the D state probability

$$\frac{\mu_d}{\mu_N} = \frac{\mu_s}{\mu_N} - \frac{3}{2}\left(\frac{\mu_s}{\mu_N} - \frac{1}{2}\right)\int_0^\infty w^2 dr; \quad \mu_N = \frac{e\hbar}{2M_p c} \qquad (16)$$

where $\mu_s = \mu_n + \mu_p$ is the isoscalar nucleon magnetic moment. $\mu_n$ and $\mu_p$ are the magnetic moments of neutron and proton and $M_p$ is the proton mass. The normalization of the wave functions require that

$$\int_0^\infty (u^2 + w^2) dr = 1 \qquad (17)$$

If we use the nondimensional parameters

$$x = (mc/\hbar)r ; u_o = u(\hbar/mc)^{1/2} ; w_o = w(\hbar/mc)^{1/2} ; Q_o = Q_d(mc/\hbar)^2$$

and E' is the binding energy, equal to 2.226 MeV, the mass of the exchange particle $m = 300$ MeV/c² (as it is evaluated from experimental data), $M_p = 939{,}573$ MeV/c², the above equations may be written as follow

$$0.319 \frac{d^2 u_o}{dx^2} - [v_c(x) + 0.00742] u_o = \sqrt{8} v_T(x) w_o$$

$$0.319 \left[ \frac{d^2 w_o}{dx^2} - \frac{6 w_o}{x^2} \right] - [v_c(x) - 2v_T(x) + 0.00742] w_o = \sqrt{8} v_T(x) u_o \qquad (18)$$

We have substituted $m/M_p = 0.319$ and $E'/mc^2 = -0.00742$. By using the experimental value of $Q_d = 0.286 \times 10^{-30}$ m², relation (15) my be rewritten as

$$\frac{1}{\sqrt{50}} \int_0^\infty x^2 u_o w_o dx - \frac{1}{20} \int_0^\infty x^2 w_o^2 dx = 0.591 \qquad (19)$$

By using experimental values of $\mu_d = 0.857 \mu_N$, $\mu_s = 0.879 \mu_N$, relation (16) becomes

$$0.857 = 0.879 - 0.569 \int_0^\infty w_o^2 dx \qquad (20)$$

Further, relation (14) becomes

$$x_m^2 = \frac{1}{4} \int_0^\infty [u_o^2 + w_o^2] x^2 dx \qquad (21)$$

and relation (17) becomes

$$\int_0^\infty (u_o^2 + w_o^2) dx = 1 \qquad (22)$$

By using the system of two coupled equations (18) and the equations (20,22) one obtains the radial functions $u_o(x)$ and $w_o(x)$ as a function of $x=(mc/\hbar)r$ where for our choice of the mass of the exchange particle $(mc/\hbar) = 1.52 \times 10^{15}$ m⁻¹.

The deuteron reduced radial wave functions $u(r)$ (upper line) and $w(r)$ (lower line) are presented in Fig.2

## 5. Conclusions

In conclusion, by using a Hamiltonian based on the coupling through flux lines we have studied the energy of interaction between two fermions via massless bosons and massive particles. In the case of the massless bosons we have found an equivalent expression of the Coulomb energy of interaction. In the case of the massive particles, we have found a potential well which is in good agreement with the experimental data for deuteron. The interaction energy contains three terms: a kinetic energy, a repulsive Coulomb energy and a nuclear potential energy. The nuclear potential well has a depth of $-0.018644131520 mc^2$ at $r = 2.18$ $(\hbar/mc)$. The contribution of the repulsive Coulomb energy increases the potential depth by $\sim 0.003 mc^2$ and displaces the minimum at $r = 2.22$ $(\hbar/mc)$ $\sim 1.4$ fm. We calculated the potential energy for the proton-neutron interaction, by taking into account the spin-spin interaction, and found that this energy is modulated through a cosine factor, whose argument contains the difference between the magnetic moments of the two particles. The difference between the energy states for the parallel and anti-parallel spins of two nucleons is of $\sim 90$ eV. In the case of the proton-neutron interaction, the proton emits a $u$ – quark and absorbs a $d$ – quark, while the neutron emits a $d$ – quark and absorbs a $u$ – quark. The two nucleons change permanently their positions.

**Appendix A**

The electron will always carry with it a lattice polarization field. The composite particle, electron plus phonon field, is called a polaron; it has a larger effective mass than the electron in the unperturbed lattice. By analogy, in our model, we consider a coupling between an electron and a boson.

Let us consider a linear chain of N bodies, separated at a distance $r$. The Hamiltonian operator of the interacting bodies ( electrons ) and the boson connecting field takes the general form

$$H = H_{o,el} + H_{o,ph} + H_I$$

where

$$H_{o,el} = \sum_{k,\sigma} \frac{\hbar^2 k^2}{2m} c^+_{k,\sigma} c_{k,\sigma}$$

is the Hamiltonian of the electrons of mass $m$, $c^+_{k,\sigma}, c_{k,\sigma}$ are the electron creation and annihilation operators, **k** is the wave vector of an electron and σ ls the spin quantum number,

$$H_{o,ph} = \sum_q \hbar \omega_q \left( a^+_q a_q + \frac{1}{2} \right)$$

$$\omega_q = \left( \frac{\alpha + Drq^2}{\rho} \right)^{1/2}$$

where $\omega_q$ is the classical oscillation frequency, $\alpha$ is the restoring force constant, $D$ is the coupling constant, $\rho$ is the linear density of flux lines, $a_q^+$, $a_q$ are the boson creation and annihilation operators and $\rho = \rho_o + Dr/c^2$, $\rho_o$ is the density of the interacting field, if this is a massive field, $c$ is the velocity of the boson waves. The interaction Hamiltonian operator $H_I$ is given by the expression

$$H_I = \frac{Dr}{2} \int \sum_n s_n \left( \frac{\partial u}{\partial z} \right)^2 \Psi^+(z) \Psi(z) dz$$

where

$$u(z) = \frac{1}{\sqrt{Nr}} \sum_{q.l} e_{ql} \left( \frac{\hbar}{2\rho \omega_{ql}} \right)^{1/2} \left( a_{ql} e^{iq.z} + a^+_{ql} e^{-iq.z} \right)$$

$$\Psi(z) = \frac{1}{\sqrt{Nr}} \sum_{k,\sigma} c_{k\sigma} e^{ik\cdot z} \chi(\sigma)$$

$$\Psi^+(z) = \frac{1}{\sqrt{Nr}} \sum_{k,\sigma} c_{k\sigma}^+ e^{-ik\cdot z} \chi(\sigma)$$

$$s_n = \frac{1}{N} \sum_{q,q_o} S_q e^{iq_o(z-z_n)}; \quad S_q = \frac{1}{r} \frac{\hbar}{2m\omega_q} \left(b_{q_o} + b_{-q_o}^+\right)$$

$b_{q_o}^+, b_{q_o}$ are creation and annihilation operators associated with the electron oscillations, $e_{ql}$ denotes the polarization vectors and $\chi(\sigma)$ is the spin wave function. $s_n$ is the displacement of a body near its equilibrium position in the direction of $r$ and, in the approximation of nearest neighbours, it is assumed that $D$ does not depend on $n$. The Hamiltonian of interaction between electrons via bosons becomes

$$H_I = -\frac{1}{N^3} \frac{D}{2r^2} \int \sum_n \sum_{k,k',q,q',q_o,\sigma,\sigma'} \frac{\hbar}{2m\omega_{q_o}} \left(b_{q_o} + b_{-q_o}^+\right) \frac{\hbar q \cdot q'}{2\rho\omega_q} ¿$$
$$e^{iq_o \cdot z_n} \left(a_q + a_{-q}^+\right)\left(a_{q'} + a_{-q'}^+\right) c_{k'\sigma'}^+ c_{k\sigma} e^{-i(q+q'+k-k'+q_o)\cdot z} dz$$

**q, q'** are the wave vectors associated with the bosons of the connecting field, **q**$_o$ is the wave vector associated with the oscillations of the electron, and **k, k'** are the wave vectors of the electrons. Consider the integral over z

$$\int e^{-i(q+q'+k-k'+q_o)\cdot z} dz = Nr\, \Delta(q+q'+k-k'+q_o) \qquad (A1)$$

where $\Delta(x) = 1$ for $x = 0$ and $\Delta(x) = 0$, otherwise. In the bulk crystal Nr is replaced by $V = N\Omega$ where $\Omega$ is the volume of a unit cell and N is the number of unit cells. We write

$$H_I = -\hbar \sum_{k,k',q,q',q_o,\sigma,\sigma'} g_{q_o}\left(a_q + a_{-q}^+\right)\left(a_{-q'} + a_{q'}^+\right) c_{k'\sigma'}^+ c_{k\sigma} b_{q_o} \Delta(q+q'+k+k'-q_o) +$$
$$g_{q_o}^¿ \left(a_{-q} + a_q^+\right)\left(b_{q_o} + b_{-q_o}^+\right) c_{k'\sigma'}^+ c_{k\sigma} b_{q_o}^+ \Delta(q+q'+k-k'+q_o)$$

where

$$g_{q_o} = \frac{\hbar D}{8N^2 mr\left(\rho_o + Dr/c^2\right)} \frac{qq'}{\omega_q^2} \sum_n e^{iq_o \cdot z_n}$$

If we omit the terms with two creators and two annihilators, it may be written

$$\left(a_q + a_{-q}^+\right)\left(a_{-q'} + a_{q'}^+\right) \approx a_q a_{q'}^+ + a_{-q}^+ a_{-q'}$$

In equation (A1) we choose **q'** $= q_o$, **k'** $= k + q$. In the interaction picture the effective Hamiltonian is given by

$$H_I^{eff} = H_{I1}^{eff} + H_{I2}^{eff}$$

$$H_{I1}^{eff} = \hbar \sum_{q,q_o,k} |g_{q_o}|^2 \frac{\omega_q}{(\varepsilon_k - \varepsilon_{k-q})^2 - \omega_q^2}$$

$$\left(a_q a_{q_o}^+ a_{q_o} a_q^+ + a_q^+ a_{q_o} a_{q_o}^+ a_q\right) c_{k-q,\sigma}^+ c_{k'\sigma'}^+ c_{k\sigma} c_{k'-q,\sigma'}$$

$$H_{I2}^{eff} = 2\hbar \sum_{q,k} |g_q|^2 \frac{1}{(\varepsilon_k - \varepsilon_{k-q}) - \omega_q} a_q^+ a_q c_{k-q,\sigma}^+ c_{k-q,\sigma}$$

The expectation value of the energy of $H_{I1}^{eff}$ is the energy of the electron-electron interaction given by equation (1) in the text. The expectation value of the energy of $H_{I2}^{eff}$ is the self energy of the electron and is used to calculate, for example, the polaron energy[3,4].

## Appendix B

TABLE I. Deuteron interaction energy, $E_{int}$ – in absence of the spin-spin interaction, $E_{int}(\Gamma_o)$ in the presence of the spin-spin interaction when the spins of the two nucleons are antiparallel, or parallel $E_{int}(\Gamma'_o)$ and nuclear potential energy, $V_N$. The energy is expressed in $mc^2$ units.

| x | $E_{int}/mc^2$ | $E_{int}(\Gamma_o)/mc^2$ | $E_{int}(\Gamma'_o)/mc^2$ | $V_N/mc^2$ |
|---|---|---|---|---|
| 0.1 | 0.078204906665 | 0.076655448098 | 0.078151018223 | 0.001504906665 |
| 0.2 | 0.044419931644 | 0.044196239933 | 0.04441220641 | 0.001719931664 |
| 0.3 | 0.032764357326 | 0.032690829341 | 0.032761821332 | 0.001397690657 |
| 0.4 | 0.026423806831 | 0.026390622907 | 0.026422662829 | 0.000723806831 |
| 0.5 | 0.022501906822 | 0.022080537370 | 0.022097488078 | 0.000201906823 |
| 0.6 | 0.018720331335 | 0.018710185049 | 0.018719981661 | -0.001313001998 |
| 0.7 | 0.015854333862 | 0.015848172309 | 0.015854121529 | -0.002559951853 |
| 0.72 | 0.01533261733 | 0.015215767966 | 0.015322219880 | -0.002812182711 |
| 0.74 | 0.014801526650 | 0.014796439827 | 0.014801393660 | -0.003087662530 |
| 0.76 | 0.014290797849 | 0.014286171429 | 0.014290638423 | -0.003356670502 |
| 0.78 | 0.013789456325 | 0.013785246876 | 0.013789311269 | -0.003628492393 |
| 0.80 | 0.013296825622 | 0.013292994586 | 0.013296693667 | -0.003903174378 |
| 0.82 | 0.012812316209 | 0.012808829255 | 0.012837211336 | -0.004180366718 |
| 0.84 | 0.012335408583 | 0.012332235058 | 0.012335299227 | -0.004459829512 |
| 0.86 | 0.011865649353 | 0.011805176760 | 0.011865549852 | -0.004741327391 |
| 0.88 | 0.011402640585 | 0.011400014380 | 0.011402560090 | -0.005034632142 |
| 0.90 | 0.010946034543 | 0.010943647503 | 0.010945952290 | -0.005309521012 |
| 0.92 | 0.010415527775 | 0.010493359895 | 0.010495453074 | -0.005595776673 |
| 0.94 | 0.010050856103 | 0.010048889297 | 0.010050788331 | -0.005883186450 |
| 0.96 | 0.009611790299 | 0.009610008181 | 0.009611727140 | --0.006171543034 |
| 0.98 | 0.009178132336 | 0.009129609698 | 0.009178076779 | -0.006460643175 |
| 1.00 | 0.008749750054 | 0.008748256023 | 0.008749661834 | -0.006750249946 |
| 1.02 | 0.008336384240 | 0.008325862857 | 0.008326339749 | -0.007040282427 |
| 1.04 | 0.007908026115 | 0.007906846713 | 0.007907985476 | -0.007330435434 |
| 1.06 | 0.007494533322 | 0.007493478021 | 0.007494498610 | -0.007620561017 |
| 1.08 | 0.007086835317 | 0.007084882449 | 0.007081939516 | -0.007910460980 |
| 1.10 | 0.006681832496 | 0.006680592663 | 0.006626112343 | -0.008199985776 |
| 1.12 | 0.006282499492 | 0.006281756112 | 0.006282473877 | -0.008488929080 |
| 1.14 | 0.005887787902 | 0.0058871333572 | 0.005887765356 | -0.008777124379 |
| 1.16 | 0.005497670003 | 0.005497097993 | 0.005497650294 | -0.009064398962 |
| 1.18 | 0.005112129304 | 0.005111633400 | 0.0051121102217 | -0.009350581560 |
| 1.20 | 0.004731159183 | 0.004730733849 | 0.004731144721 | -0.009635507284 |
| 1.22 | 0.004354763113 | 0.004354402646 | 0.004374546413 | -0.009919007379 |
| 1.24 | 0.003982951865 | 0.003982651557 | 0.003982941518 | -0.010200919103 |
| 1.26 | 0.003615744822 | 0.003615500125 | 0.003615736391 | -0.010481080575 |
| 1.28 | 0.003253168398 | 0.003252975094 | 0.003253166737 | -0.010759331602 |
| 1.30 | 0.002895295698 | 0.002895109772 | 0.002895256574 | -0.011035513632 |
| 1.32 | 0.002542045662 | 0.002541943664 | 0.002542042138 | -0.011309469500 |
| 1.34 | 0.002193583568 | 0.002193522035 | 0.002193581448 | -0.011581043297 |
| 1.36 | 0.001849919712 | 0.001849895485 | 0.001849918878 | -0.011850080288 |
| 1.38 | 0.001511109568 | 0.001511119711 | 0.001511109918 | -0.012116426684 |
| 1.40 | 0.001177213381 | 0.001177255160 | 0.001177214078 | -0.012379929476 |
| 1.42 | 0.000848296017 | 0.00084836680 | 0.000848293459 | -0.012640436377 |
| 1.44 | 0.000524426622 | 0.000524524156 | 0.000524429984 | -0.12897795600 |
| 1.46 | 0.000205678742 | 0.000205800782 | 0.000205682947 | -0.013151855505 |
| 1.48 | -0.000107870430 | -0.000107725995 | -0.00107865454 | -0.013402465026 |
| 1.50 | -0.000416139139 | -0.000415974252 | -0.000416133457 | -0.013649473473 |
| 1.52 | -0.000719042070 | -0.000718858552 | -0.000719035747 | -0.013892726281 |
| 1.54 | -0.001016371520 | -0.001016289454 | -0.001016482991 | -0.014131955936 |
| 1.56 | -0.001308387124 | -0.001326829656 | -0.001308379669 | -0.014367361482 |
| 1.58 | -0.001594643371 | -0.001594413762 | -0.001453034279 | -0.014598440839 |
| 1.60 | -0.001875150191 | -0.001874906404 | -0.001875141851 | -0.014825150191 |
| 1.62 | -0.002149805893 | -0.002149552736 | -0.002149797220 | -0.015047336757 |
| 1.64 | -0.002418498953 | -0.002418231913 | -0.002418489890 | -0.015264840417 |
| 1.66 | -0.002681118624 | --0.002680846858 | -0.002681109260 | -0.015477504166 |
| 1.68 | -0.002937543021 | -0.002937267617 | -0.002937533394 | -0.015685162069 |
| 1.70 | -0.003187660002 | --0.003187364003 | -0.003187640147 | -0.015887650002 |
| 1.72 | -0.003431310797 | -0.003431019118 | -0.003434300747 | -0.016084799169 |
| 1.74 | -0.003668391114 | -0.003668094675 | -0.003658380900 | -0.016276437091 |
| 1.76 | -0.003896750749 | -0.003898450402 | -0.003898740400 | -0.016452387113 |
| 1.78 | -0.00412243166 | -0.004121939715 | -0.004122232721 | -0.016642467896 |
| 1.80 | -0.004400274668 | -0.004338408840 | -0.0043387004131 | -0.016816492446 |
| 1.90 | -0.005309803992 | -0.005309456092 | -0.005309794069 | -0.017588751360 |
| 1.92 | -0.005480439473 | -0.005480132854 | -0.005489428908 | -0.017722106139 |
| 1.94 | -0.005642715548 | -0.005642410643 | -0.005642705044 | -0.017847870187 |
| 1.96 | -0.005796389733 | -0.005796084936 | -0.00057963799306 | -0.01765777489 |
| 1.98 | -0.005941201271 | -0.005940901324 | -0.005941190877 | -0,018075544706 |
| 2.00 | -0.006076869209 | -0.006076572341 | -0.006076858980 | -0.018176869209 |
| 2.02 | -0.006203089170 | -0.006202796645 | -0.006203079061 | -0.018269425804 |
| 2.04 | -0.006319530497 | -0.006319246934 | -0.006319520520 | -0.018352863831 |
| 2.06 | -0.006425830575 | -0.006425545197 | -0.006425820742 | -0.018427607733 |
| 2.08 | -0.006521591284 | -0.006521310427 | -0.006521581125 | -0.018490822055 |
| 2.10 | -0.006606371569 | -0.006606095539 | -0.006606362058 | -0.018544466807 |
| 2.12 | -0.006679679899 | -0.006679409004 | -0.006679670565 | -0.018587227069 |
| 2.14 | --0.00674096400 | -0.006740698540 | -0.006740954862 | -0.18618934103 |
| 2.16 | -0.006789598007 | -0.006789338242 | -0.006789589057 | -0.018637746155 |
| 2.18 | -0,006824865465 | -0.006796295859 | -0.006824611678 | *-0.018644131520* |
| 2.20 | -0.0068459364408 | -0.006845689379 | -0.006845928393 | -0.018636846013 |
| 2.22 | *-0.006851838986* | *-0.006851517954* | *-0.006851830681* | -0.018614902049 |
| 2.24 | -0.006841412034 | -.0.006841177779 | -0.006841403963 | -0.018577126320 |
| 2.26 | -0.006813247344 | -0.006813020140 | -0.006813235516 | -0.018522096902 |
| 2.28 | -0.006765594820 | -0.006765384830 | -0.006765587244 | -0.018448050960 |
| 2.30 | -0.006690185596 | -0.006696004974 | -0.006696208883 | -0.018352737934 |
| 2.32 | -0.006602140610 | -0.006601935674 | -0.006602132867 | -0.018233174384 |
| 2.34 | -0.006479216175 | -0.006479020358 | -0.006479209427 | -0.018085199081 |
| 2.36 | -0.006321212424 | -0.006321025478 | -0.006321205981 | -0.017902568356 |
| 2.38 | -0.006117631166 | -0.006112453793 | -0.006117625054 | -0.0176744023 |
| 2.40 | -0.005846558802 | -0.005846392088 | -0.005846553058 | -0.017379852136 |
| 2.42 | -0.005406109332 | -0.005405958626 | -0.005406106267 | -0.016916026688 |
| 2.45 | -0.004801022851 | -0.004800888303 | -0.004801018222 | -0.016916026688 |
| 2.5 | -0.004220704459 | -0.004220589564 | -0.004220700500 | -0.015640704459 |
| 2.6 | -0.003478146575 | -0.003478058010 | -0.003478143523 | -0.014793531190 |
| 2.7 | -0.002979046775 | -0.002978976007 | -0.002979044339 | -0.014197566294 |
| 2.8 | -0.002608266231 | -0.002608208497 | -0.002608264219 | -0.013736837600 |
| 2.9 | -0.002318480760 | -.0.002318432785 | -0.002318479107 | -0.013363308346 |
| 3.0 | -0.002135895260 | -0.002084554926 | -0.001913593862 | -0.013051261918 |
| 3.1 | -0.001891476800 | -0.001891442551 | -0.001891475620 | -0.012785025187 |
| 3.2 | -0.001729242943 | -0.001725322588 | -0.001729241932 | -0.012554342943 |
| 3.3 | -0.001591076024 | -0.001591050468 | -0.001591074942 | -0.012351682083 |
| 3.4 | -0.001472977826 | -0.001472046996 | -0.001472077066 | -0.012172077826 |
| 3.5 | -0.00123683886677 | -0.001368800138 | -0.001368818763 | -0.012011477967 |
| 3.6 | -0.001277943159 | -0.001277926087 | -0.001277942476 | -0.011866831950 |
| 3.7 | -0.001197910492 | -0.001197895479 | -0.001197909975 | -0.011735748330 |
| 3.8 | -0.0011268337904 | -0.001126821820 | -0.001126834701 | -0.011616311588 |
| 3.9 | -0.00106366048 | -0.001063355368 | -0.001063366864 | -0.011506955791 |
| 4.1 | -0.000955075794 | -0.000955066034 | -0.000955075300 | -0.011313612220 |
| 4.2 | -0.000908613792 | -0.000908605129 | -0.000908613493 | -0.011227666411 |
| 4.3 | -0.000866408343 | -0.000866400499 | -0.000866408072 | -0.011147803612 |
| 4.4 | -0.000827938580 | -0.000827931454 | -0.000827938334 | -0.011073393125 |
| 4.5 | -0.000792763412 | -0.000792756920 | -0.000798763188 | -0.011003874523 |
| 4.6 | -0.000760506786 | -0.000760500854 | -0.000760506581 | -0.010938767655 |
| 4.7 | -0.000730846202 | -0.000730840770 | -0.000730846015 | -0.010877654710 |
| 4.8 | -0.000703503603 | -0.000703498616 | -0.000703503431 | -0.010820170270 |
| 4.9 | -0.0006778238040 | -0.000678233450 | -0.000678237882 | -0.010766074914 |
| 5.0 | -0.000654839761 | -0.00065483852 | -0.000654839615 | -0.010714839761 |
| 5.1 | -0.000633125385 | -0.000633118446 | -0.000633125250 | -0.010666468718 |
| 5.2 | -0.000612933939 | -0.000612930331 | -0.000612933814 | -0.010620626247 |
| 5.3 | -0.000594123596 | -0.000594120248 | -0.000594123480 | -0.010577142464 |
| 5.4 | -0.000576568951 | -0.00576665838 | -0.000576568843 | -0.010535828210 |
| 5.5 | -0.000560158757 | -0.000560155858 | -0.000560158656 | -0.010496522393 |
| 5.6 | -0.000544793757 | -0.000544791311 | -0,00054538976 | -0.010459079471 |
| 6.0 | -0.000492154795 | -0.000492152714 | -0.000492154722 | -0.010325488128 |
| 6.5 | -0.000441572310 | -0.000441570764 | -0.000441572256 | -0.010187726156 |
| 7.0 | -0.000402879758 | -0.000402878567 | -0.000402879717 | -0.010074308329 |
| 7.5 | -0.000372593929 | -0.000372595002 | -0.000372593897 | -0.009979260596 |
| 8.0 | -0.000348427445 | -0.00348426703 | -0.000348427419 | --0.009898427665 |
| 8.5 | -0.000328825344 | -0.000328824739 | -0.000328825323 | -0.009828825344 |
| 9.0 | -0.000312699375 | -0.000312698874 | -0.000312699358 | -0.009768254931 |
| 9.5 | -0.000299268762 | -0.000299268342 | -0.000299268748 | -0.009715058236 |
| 10 | -0.000287960402 | -0.000287960806 | -0.000287961150 | -0.009667960022 |
| 11 | -0.00270107774 | -0.000270107508 | -0.000270107766 | -0.009588289593 |
| 12 | -0.000256791475 | -0.000256791270 | -0.000256791468 | -0.009523458142 |
| 13 | -0.000246588808 | -0.000246588645 | - 0.000246588802 | -0.009469665731 |
| 14 | -0.000238595296 | -0.000238595012 | -0.0002328595143 | -0.009424309432 |
| 15 | -0.000232212859 | -0.000232212750 | -0.000232212856 | -0.009385546193 |

**Figure Captions**

Fig.1 The reduced interaction energy between a proton and a neutron via mass less bosons and massive particles, as a function of the reduced distance between them. The lower line curve is the potential nuclear interaction $v = V_N/mc^2$ while the upper line curve is the total energs of interaction $E = E_{int}/mc^2$ , while, the sinuous curve is in presence of the spin-spin interaction.

Fig.2. The deuteron reduced radial wave functions $u_o$ ( upper line) and $w_o$ (lower line) for our potential of interaction, as a function of reduced relative coordinate.

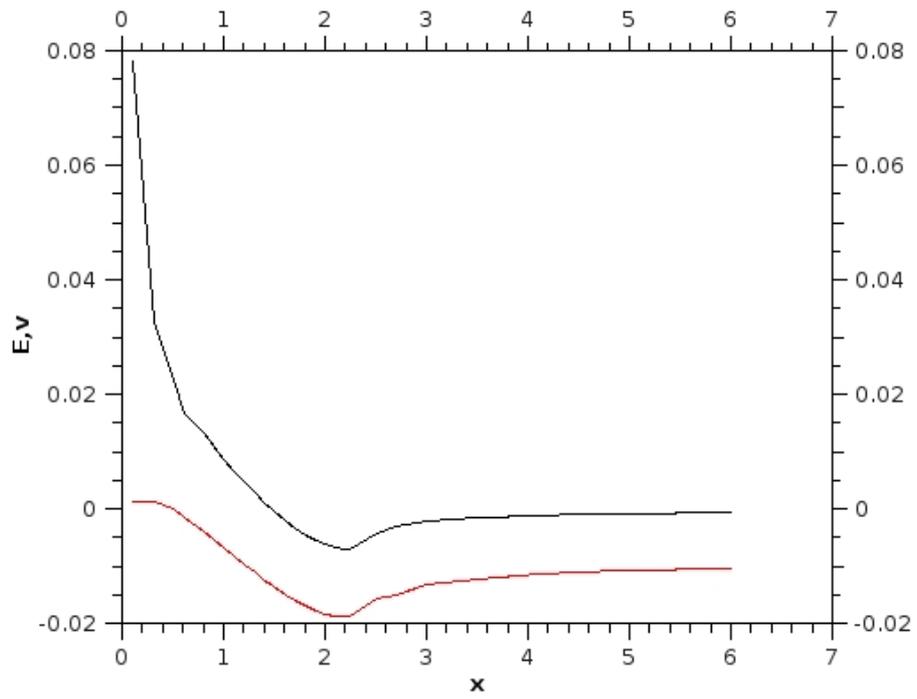

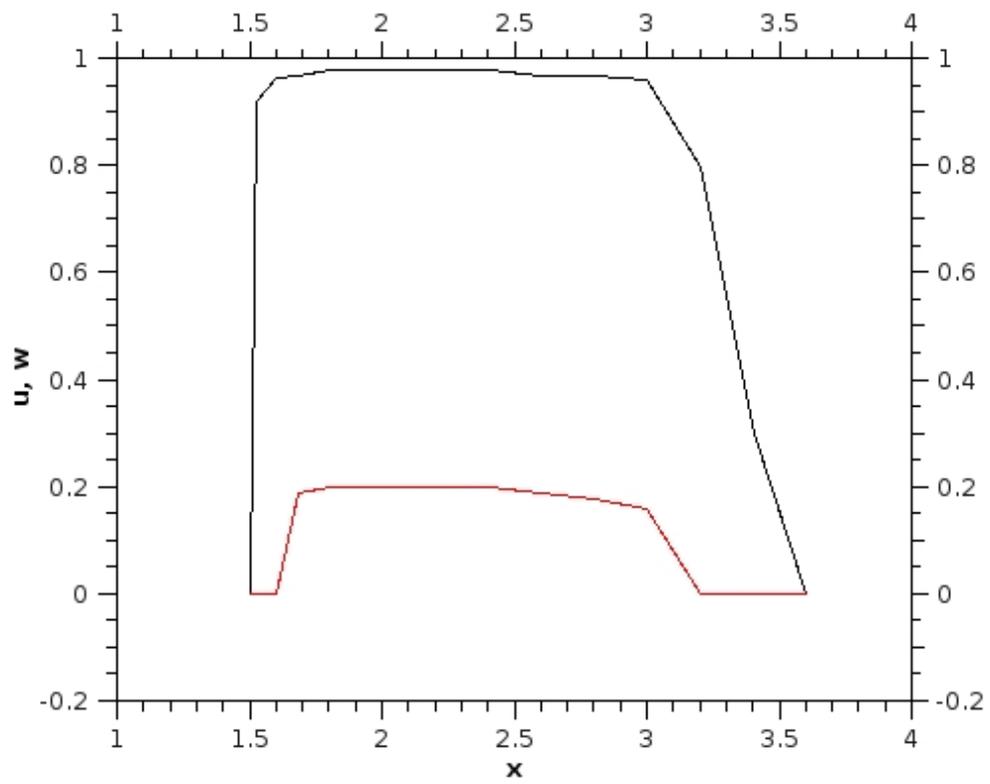